\documentclass{article}
\usepackage{spconf,amsmath,graphicx,hyperref}
\usepackage{bm}
\usepackage{amssymb}
\usepackage{multirow}
\usepackage{multicol}
\usepackage{booktabs}
\usepackage{caption}
\usepackage{subfig}
\usepackage{cite}
\usepackage{color}
\usepackage{mathtools}

\DeclareMathOperator*{\argmin}{arg\,min}

\DeclareMathOperator*{\diag}{blkdiag}
\DeclareMathOperator*{\concat}{cat}

\title{Time-varying rPPG signal separation via block-sparse signal model}
\name{
Kosuke Kurihara$^{\star}$,~
Yoshihiro Maeda$^{\dagger}$,~
Daisuke Sugimura$^{\ddagger}$,~and~
Takayuki Hamamoto$^{\star}$
\thanks{This work was supported by JSPS KAKENHI JP24K23903, JP25K21231, and JP24K02964.}
\thanks{This article has been accepted for publication in IEEE International Conference on Image Processing (ICIP) 2026. This is the author’s version which has not been fully edited and content may change prior to final publication.}
\thanks{
\copyright 2026 IEEE. Personal use of this material is permitted. Permission
from IEEE must be obtained for all other uses, in any current or future media,
including reprinting/republishing this material for advertising or promotional
purposes, creating new collective works, for resale or redistribution to servers
or lists, or reuse of any copyrighted component of this work in other works.}}
\address{$^{\star}$Tokyo University of Science \, $^{\dagger}$Shibaura Institute of Technology \, $^{\ddagger}$Tokyo Metropolitan University}
\begin{document}
\maketitle
\begin{abstract}
Remote photoplethysmography (rPPG) enables non-contact measurement of cardiac pulse signals by analyzing subtle color changes in facial videos.
Nevertheless, extracting rPPG signals remains challenging because of their extremely weak signal strength and susceptibility to illumination noise.
In this paper, we propose an rPPG signal extraction method that exploits the quasi-periodic characteristics of rPPG signals.
Our approach models quasi-periodicity of the rPPG signal, which arises from the stable cardiac cycle, as a block-sparse structure in the time-frequency domain.
To incorporate a block-sparse model and enable adaptive signal separation under illumination fluctuations, we construct a time-varying signal separation framework.
Experiments using a public dataset demonstrate the effectiveness of our method.
\end{abstract}
\section{Introduction}
\label{sec:intro}
Photoplethysmography (PPG) signals are time-series signals that represent blood volume changes associated with cardiac cycles~\cite{PPG_Application}.
Traditionally, contact-type sensors have been used as the gold standard for PPG signal measurement; however, these sensors require continuous physical contact during measurement, imposing constraints on daily usage scenarios.

Camera-based PPG measurement, known as remote PPG (rPPG), has emerged as a promising approach for continuous vital sign monitoring without physical contact~\cite{survey_2024}.
By analyzing subtle variations in the reflected light from the skin region, rPPG signals can be extracted.

The fundamental challenge of rPPG signal extraction lies in weak signal strength and illumination noise, which are inherent problems due to the non-contact nature of the rPPG technique.
According to a previous study~\cite{DistancePPG}, the amplitude of the rPPG components is quite small, less than 2 bits in typical 8-bit depth RGB cameras.
Thus, even under bright illumination conditions, the rPPG signal is obscured by fluctuations in ambient illumination and other noise sources.

To overcome these challenges, many rPPG signal extraction methods have been proposed.
Verkruysse \textit{et al.}~\cite{remoteAmbient} analyzed the influence of RGB channel selection on rPPG signal extraction.
They analyzed RGB signals by applying a band-pass filter and frequency analysis, and demonstrated that the green channel is the most reliable for rPPG signal extraction.

Several rPPG signal extraction methods exploited the physiological characteristics of skin and blood~\cite{POS,CHROM}.
Haan \textit{et al.}~\cite{CHROM} proposed a chrominance analysis method based on a dichromatic reflection model.
By analyzing color changes in their chrominance space, they extracted rPPG-derived color variations while suppressing luminance variations caused by illumination fluctuations.
Although chrominance approaches improved the rPPG signal extraction performance, temporal variation in the illumination color spectrum deteriorates the chrominance space, leading to performance degradation.

The temporal characteristics of rPPG signals have been utilized~\cite{PohICA,macwanPVM,Macwan2018}.
Poh \textit{et al.}~\cite{PohICA} proposed an rPPG signal extraction method based on a blind source separation framework.
They assumed that the rPPG signal and other noise components are statistically independent, and then extracted the rPPG signal by using independent component analysis (ICA).
The methods~\cite{macwanPVM,Macwan2018} utilized the periodic characteristics of the rPPG signals.
Based on the knowledge that rPPG signals have temporal periodicity derived from cardiac cycles~\cite{ppgperiod}, they modeled the rPPG signals as having high temporal autocorrelation.
They then extracted the rPPG signal by finding a separation vector that maximizes its autocorrelation.

In spite of these advances, temporal modeling based on independence or autocorrelation may be insufficient.
Independence-based methods~\cite{PohICA} only assume statistical independence without considering the unique characteristics of rPPG; thus, these methods may exhibit limited performance when illumination fluctuations dominate over weak rPPG components.
Although autocorrelation-based methods~\cite{macwanPVM,Macwan2018} consider the temporal periodicity of the rPPG signal, they only aim to maximize autocorrelation in the time domain and do not consider the unique time-frequency structure of the rPPG signals.

In this paper, we propose an rPPG signal extraction method that exploits the unique temporal characteristics of rPPG signals.
The main contribution is the fine-grained modeling of rPPG signals through the block-sparse model in the time-frequency domain.
rPPG signals reflect the regularity of cardiac cycles, giving rise to quasi-periodic oscillations with stable dominant frequency components across consecutive time windows.
By representing this unique rPPG characteristic with a block-sparse model, we can capture the quasi-periodic structure of rPPG signals more precisely.
Another contribution lies in the construction of a time-varying separation framework that integrates our block-sparse model.
By formulating the rPPG signal separation for each short time window, we can adapt to fluctuating illumination while incorporating the block-sparse model of rPPG signals in the time-frequency domain.

\vspace{-1.5mm}
\section{Proposed method}
\vspace{-1.5mm}
\label{sec:prop}
Our method consists mainly of three steps after preprocessing: (1) constructing a time-varying signal separation framework, (2) formulating an objective function with our block-sparse model, and (3) optimizing via alternating minimization.
In the following, we describe each step in detail.

\vspace{-1mm}
\subsection{Preprocessing}
\vspace{-1mm}
\label{sec:preprocessing}
We first obtain input RGB time-series signals from a patch defined in the face area for each frame.
By averaging the pixel values within the patch, we obtain RGB signals.
We denote the input RGB signals as $\widetilde{\bm{X}} \in \mathbb{R}^{L \times C}$, where $L$ and $C$ respectively denote the number of frames and channels in the RGB video (i.e., $C=3$).

To realize time-varying rPPG separation, we apply a sliding window to $\widetilde{\bm{X}}$ with stride one, obtaining $T$ windowed signals.
The $t\in\{1,\ldots,T\}$-th windowed signal of $\widetilde{\bm{X}}$ is denoted by $\bm{X}_t \in \mathbb{R}^{\tau \times C}$, where $\tau$ denotes the window size.

\vspace{-1mm}
\subsection{Time-varying signal separation}
\vspace{-1mm}
\label{sec:separation}
We construct a time-varying rPPG signal separation framework.
Previous studies have shown that RGB linear mixture models are effective for rPPG signal extraction~\cite{POS,PohICA}.
However, these methods employ a static separation vector, which cannot adapt to fluctuating illumination and is incompatible with our block-sparse model in the time-frequency domain.
We therefore extend the above-mentioned linear mixture model by reformulating it for each short time window.

The $t$-th windowed rPPG signal $\bm{y}_t \in \mathbb{R}^{\tau} $ can be formulated by a linear combination of the input RGB signal $\bm{X}_t$ as
\begin{equation}
    \label{eq:separation}
    \bm{y}_t = \bm{X}_t \bm{w}_t + \bm{e}_t ~,
\end{equation}
where $\bm{w}_t \in \mathbb{R}^{C}$ and $\bm{e}_t \in \mathbb{R}^{\tau}$ denote the $t$-th separation vector and the residual error, respectively.

We now formulate the rPPG signal separation problem over all time windows.
Let $\concat(\cdot, \ldots, \cdot)$ denote the concatenation operator that vertically stacks input vectors.
Based on Eq.~\eqref{eq:separation}, the concatenation of the windowed rPPG signals, defined as ${\bm{y}} = \concat(\bm{y}_1, \ldots, \bm{y}_T)$, can be written as
\begin{equation}
    \label{eq:entire_separation}
    {\bm{y}} = \mathcal{X}  {\bm{w}} + \bm{e} ~,
\end{equation}
where $\mathcal{X} = \diag(\bm{X}_1, \ldots, \bm{X}_T) \in \mathbb{R}^{\tau T \times CT}$ denotes the block diagonal matrix containing $\bm{X}_t$.
In addition, ${\bm{w}} = \concat(\bm{w}_1, \ldots, \bm{w}_T) \in \mathbb{R}^{C T} $ and ${\bm{e}} =  \concat(\bm{e}_1, \ldots, \bm{e}_T) \in \mathbb{R}^{\tau T} $ denote the concatenated vector of $\bm{w}_t$ and $\bm{e}_t$, respectively.

To relate the entire rPPG signal $\widetilde{\bm{y}} \in \mathbb{R}^{L}$ and the concatenation of windowed rPPG signals $\bm{y} \in \mathbb{R}^{\tau T}$, we introduce the sliding window operator $\bm{G} \in \mathbb{R}^{\tau T \times L}$ as
\begin{equation}
    \label{eq:sliding_window_matrix}
    \bm{G} = \left[\begin{array}{lll}
            \bm{I}_{\tau} \, \bm{0}  & \ldots                           \\
            \bm{0}_{\tau} \, \bm{I}_{\tau} \, \bm{0} & \ldots                  \\
            \bm{0}_{\tau} \, \bm{0}_{\tau} \, \bm{I}_{\tau} \, \bm{0} \\
             \vdots  &        \!\!\!\!\!\!\!\!\!\!\!    \ddots &  ~ \\
        \end{array}\right] ~,
\end{equation}
where $\bm{I}_{\tau} \in \mathbb{R}^{\tau \times \tau}$ denotes the identity matrix, and $\bm{0}_{\tau} \in \mathbb{R}^{\tau}$ and $\bm{0}$ denote zero vector and zero matrix, respectively.

With $\bm{G}$, the relationship between $\widetilde{\bm{y}}$ and $\bm{y}$ is expressed as
\vspace{-1.5mm}
\begin{equation}
\label{eq:sliding_window_operator}
{\bm{y}} = \bm{G} \widetilde{\bm{y}}~.
\end{equation}

We finally obtain the time-varying signal separation framework based on Eqs.~\eqref{eq:entire_separation}, \eqref{eq:sliding_window_matrix}, and \eqref{eq:sliding_window_operator} as
\begin{equation}
    \label{eq:final_sepa_model}
    \bm{G} \widetilde{\bm{y}} = \mathcal{X} \bm{w} + \bm{e} ~.
\end{equation}

\subsection{Objective function}
\label{sec:construct}
Based on the signal separation framework in Sec.~\ref{sec:separation}, we design the objective function for the rPPG signal extraction.
Through joint estimation of the rPPG signal $\widetilde{\bm{y}}$ and the separation vector $\bm{w}$, we formulate the optimization problem as
\begin{align}
    \label{eq:objective}
    \widetilde{\bm{y}}, \bm{w} = \argmin_{\widetilde{\bm{y}}_*,  \bm{w}_* } \|\bm{G} \widetilde{\bm{y}}_*\!-\!\mathcal{X} \bm{w}_* \|_2^2
     +  f_1(\widetilde{\bm{y}}_*)  +  f_2(\bm{w}_*) \, ,
\end{align}
where $f_1(\widetilde{\bm{y}})$ and $f_2(\bm{w})$ denote prior terms that model $\widetilde{\bm{y}}$ and $\bm{w}$, respectively.
In the following, we describe the details of each term and the optimization procedure for solving Eq.~\eqref{eq:objective}.

\vspace{-1mm}
\subsection{Block-sparse model in time-frequency domain}
\label{sec:periodicity}
We describe $f_1(\widetilde{\bm{y}})$ in Eq.~\eqref{eq:objective} that models the quasi-periodicity of rPPG signals $\widetilde{\bm{y}}$ in the time-frequency domain.
To obtain the time-frequency representation of $\widetilde{\bm{y}}$, we first apply the short-time Fourier transform (STFT) by using the block diagonal matrix $\mathcal{F} = \diag(\bm{F},\ldots,\bm{F}) \in \mathbb{C}^{\tau T \times \tau T}$, where $\bm{F} \in \mathbb{C}^{\tau \times \tau}$ is the discrete Fourier transform matrix for a single time window.
By applying $\mathcal{F}$ to the concatenation of windowed rPPG signals $\bm{G}\widetilde{\bm{y}}$ introduced in Eq.~\eqref{eq:sliding_window_operator}, we obtain the time-frequency representation $\mathcal{F} \bm{G} \widetilde{\bm{y}} \in \mathbb{C}^{\tau T}$.

We then model $\mathcal{F} \bm{G} \widetilde{\bm{y}}$ to capture its structured sparsity.
Due to the regularity of cardiac cycles, rPPG signals exhibit quasi-periodicity with a few dominant frequency components~\cite{ppgperiod}, and these components remain consistent across consecutive time windows.
To promote this time-frequency structure, we introduce the block-sparse model~\cite{blocksparse}.
In this model, blocks are formed by grouping all temporal components within each frequency bin, and nonzero components are concentrated in a few blocks.

To incorporate a block-sparse model of $\mathcal{F} \bm{G} \widetilde{\bm{y}}$, we design the prior term $f_1(\widetilde{\bm{y}})$ using the weighted $\ell_{2,1}$ norm as
\vspace{-2mm}
\begin{equation}
    \label{eq:frequency_sparsity}
    f_1(\widetilde{\bm{y}}) = \| \mathcal{F}\bm{G}\widetilde{\bm{y}} \, \|_{2,1}^{\bm{\alpha}} \coloneqq \sum_{f=1}^{\tau} \alpha_f \| [ \, \mathcal{F}\bm{G}\widetilde{\bm{y}} \, ]_f \|_2 \,,
    \vspace{-1mm}
\end{equation}
where $\|\cdot\|_2$ and $\|\cdot\|_{2,1}^{\bm{\alpha}}$ denote the $\ell_2$ norm and the weighted $\ell_{2,1}$ norm with positive weights $\bm{\alpha} = [\alpha_1, \ldots, \alpha_{\tau}]^{\top} \in \mathbb{R}_{>0}^{\tau}\,$, respectively.
In addition, $[ \bm{b} ]_f \in \mathbb{C}^{T}$ denotes the $f$-th frequency block of the time-frequency representation $\bm{b} \in \mathbb{C}^{\tau T}$, obtained by concatenating the $f$-th frequency component across all time windows in $\bm{b}$.

The weight $\alpha_f$ controls the sparsity of the $f$-th frequency block.
Adjusting $\alpha_f$ according to the frequency bands in which the rPPG signals are expected to be present, we can promote sparsity based on the rPPG characteristics while suppressing noise in other frequency bands.

\vspace{-2mm}
\subsection{Prior on time-varying separation vector}
\vspace{-1mm}
\label{sec:constraint}
Although our time-varying separation vector enables adaptation to fluctuating illumination, additional priors are required to ensure stable and meaningful solutions.
To that end, we impose two priors on the separation vectors $\bm{w}$.
First, we enforce temporal smoothness to ensure that the separation vectors between adjacent time windows change gradually, preventing abrupt changes that could introduce artifacts.
Second, we impose a unit norm constraint to avoid trivial solutions such as $\bm{w}_t=\bm{0}$ induced by temporal smoothness.

Based on this, the prior term $f_2(\bm{w})$ in Eq.~\eqref{eq:objective} is given by
\begin{equation}
    \label{eq:smoothness}
    f_2(\bm{w}) = \beta \|\bm{D} \bm{w}\|_2^2 \, + \, \phi(\bm{w}) \, ,
\end{equation}
where $\beta$ denotes the weight for the first term promoting temporal smoothness of $\bm{w}$.
The matrix $\bm{D} \in \mathbb{R}^{3(T-1) \times 3T}$ represents the difference operator for adjacent separation vectors:
\begin{equation}
\bm{D} = \left[\begin{array}{rrrrrrr}
        1      & 0     & 0       &  -1      &  0 & 0&\ldots      \\
        0      & 1      & 0     & 0       &  -1    &0&\ldots   \\
               &        &  & \ddots &      &  \\
    \end{array}\right] ~.
\end{equation}
The second term $\phi(\bm{w})$ represents the indicator function that constrains unit norm, defined as
\begin{equation}
    \label{eq:indicator}
    \phi (\bm{w}) =
    \begin{cases}
        0       & \mathrm{if~~} \| \bm{w}_t \|_2 = 1 ~, ~~ \forall t \in \{ 1, \ldots, T \} \\
        +\infty & \mathrm{otherwise} \, .
    \end{cases}
\end{equation}

\subsection{Alternating minimization}
\label{sec:optimization}
Because the joint estimation of the rPPG signal $\widetilde{\bm{y}}$ and the separation vector $\bm{w}$ in Eq.~\eqref{eq:objective} is non-convex, we alternately update $\widetilde{\bm{y}}$ and $\bm{w}$.

\subsubsection{Update of rPPG signal}
We update $\widetilde{\bm{y}}$ based on the following optimization problem with $\bm{w}$ fixed:
\begin{equation}
    \label{eq:update_Y}
        \widetilde{\bm{y}} \! =
       \argmin_{ \widetilde{\bm{y}}_* }
          \|\bm{G} \widetilde{\bm{y}}_* - \mathcal{X} \bm{w} \|_2^2  +  \| \mathcal{F}\bm{G}\widetilde{\bm{y}}_* \, \|_{2,1}^{\bm{\alpha}} \,.
\end{equation}

To handle a non-differentiable weighted $\ell_{2,1}$ norm in Eq.~\eqref{eq:update_Y}, we employ the alternating direction method of multipliers (ADMM)~\cite{proximal} with the proximal operator.
Specifically, we iteratively solve the following subproblems.
\begin{align}
\widetilde{\bm{y}}^{(k)} &\,=\, \argmin_{\widetilde{\bm{y}}_*} \| \bm{G} \widetilde{\bm{y}}_* - \mathcal{X}\bm{w}\|_2^2 \nonumber \\
&~~~~~~~~~~~~~~~~~~+ \frac{1}{2\gamma} \| \mathcal{F} \bm{G} \widetilde{\bm{y}}_* - \bm{z}^{(k-1)} + \bm{v}^{(k-1)} \|_2^2 \, , \label{eq:admm_yupdate} \\
\bm{z}^{(k)} &\,=\,  \mathrm{prox}_{\gamma, \| \cdot \|_{2,1}^{\bm{\alpha}}} (\mathcal{F} \bm{G} \widetilde{\bm{y}}^{(k)} + \bm{v}^{(k-1)}) \, , \label{eq:admm_zupdate} \\
\bm{v}^{(k)} &\,=\,  \bm{v}^{(k-1)} + \mathcal{F}\bm{G}\widetilde{\bm{y}}^{(k)} -\bm{z}^{(k)} \, , \label{eq:admm_vupdate}
\end{align}
where $\gamma$ denotes the control parameter.
The vectors $\bm{z}^{(k)}$ and $\bm{v}^{(k)}$ are the auxiliary variables for the $k$-th ADMM iteration.
In addition, $\mathrm{prox}_{\gamma, \| \cdot \|_{2,1}^{\bm{\alpha}}}(\cdot)$ denotes the proximal operator for the weighted $\ell_{2,1}$ norm.

We briefly explain the procedure for solving Eqs.~\eqref{eq:admm_yupdate} and \eqref{eq:admm_zupdate}.
Since Eq.~\eqref{eq:admm_yupdate} is a quadratic problem, we solve it by using the conjugate gradient method~\cite{interiorpoint}.
For Eq.~\eqref{eq:admm_zupdate}, we apply the proximal operator $\mathrm{prox}_{\gamma, \| \cdot \|_{2,1}^{\bm{\alpha}}} (\cdot)$, which performs block-wise soft-thresholding~\cite{proximal}:
\begin{align}
    [\mathrm{prox}_{\gamma, \| \cdot \|_{2,1}^{\bm{\alpha}}} \! (\bm{u}) ]_f =
    \begin{cases}
    \!\left(1 \! - \! \frac{\gamma \alpha_f}{\| \bm{[u]}_f \|_2} \right) \! \bm{[u]}_f  &\mathrm{if} ~ \| \bm{[u]}_f \|_2 > \gamma \alpha_f \\
    \bm{0}                                       & \mathrm{otherwise}~ .
    \end{cases}
\end{align}
\vspace{-8mm}
\subsubsection{Update of separation vector}
Given the estimated $\widetilde{\bm{y}}$, we update $\bm{w}$ by solving the following optimization problem:
\begin{equation}
    \label{eq:update_W}
    \bm{w} = \argmin_{\bm{w}_*} \|\bm{G} \widetilde{\bm{y}} - \mathcal{X} \bm{w}_* \|_2^2 \, + \beta \|\bm{D} \bm{w}_*\|_2^2 \, + \, \phi(\bm{w}_*) \,.
\end{equation}

Since Eq.~\eqref{eq:update_W} can be treated as a constrained nonlinear least squares problem, we solve it using the interior point method~\cite{interiorpoint}.
We alternately update $\widetilde{\bm{y}}$ and $\bm{w}$ until convergence to obtain the final estimate of the rPPG signal.

\section{Experiments}
\label{sec:experiments}
\subsection{Experimental settings}
\subsubsection{Dataset}
We conducted experiments using the UBFC-RPPG dataset~\cite{Bobbia_PR2019_UBFC_dataset}.
This dataset comprises 49 videos of 47 subjects who were instructed to sit.
Each video was captured for approximately 1 minute.
The illumination conditions were natural and not strictly controlled.
The captured videos are in an uncompressed 8-bit RGB format with a resolution of $640\times480$ pixels and a frame rate of 30 fps.

\subsubsection{Evaluation settings}
\label{sec:evalset}
Our method provides a principled approach to rPPG signal extraction from a single patch; in contrast, recent learning-based methods improve performance by leveraging multiple patches.
To enable comparison with these methods, we conducted experiments in both single and multiple patch settings.

Following~\cite{ourtip,ouraccess}, we first obtained a set of patches from the face area.
For each method, we extracted the rPPG signal from each patch.
In the single patch setting, we evaluated individual performance using each obtained rPPG signal.
In the multiple patch setting, we further aggregated the rPPG signals obtained from each patch by using principal component analysis (PCA), a common technique to leverage multiple patch information in rPPG extraction~\cite{selfMatrix}.
We then evaluated the performance using the aggregated signal.

\subsubsection{Comparison methods}
For comparison, we used \textit{Green}~\cite{remoteAmbient}, which performs band-pass filtering on green signals.
For chrominance-based methods, we used two methods: \textit{CHROM}~\cite{CHROM} and \textit{POS}~\cite{POS}.
In addition, we employed \textit{ICA}~\cite{PohICA} and \textit{PVM}~\cite{macwanPVM}, which are ICA and autocorrelation based methods, respectively.

In the multiple patch setting, we additionally compared our extended method with \textit{MTTS-CAN}~\cite{mttscan} and \textit{PhysFormer}~\cite{PhysFormer}, which are CNN and transformer-based methods, respectively.
We used the pre-trained models provided by these authors.
We also compared with multiple patch variants of single patch based methods~\cite{remoteAmbient,CHROM,POS,PohICA,macwanPVM} by extending them with PCA.
In single and multiple patch settings, we used the same set of patches for all methods for fair comparison.

\begin{table}[t]
    \centering
    \caption{Performance comparison for single patch setting. The best results are highlighted in \textbf{bold}.}
    \label{tab:results}
\begin{tabular}{l|ccc} \toprule
    \multirow{2}{*}{Method} & SNR $\uparrow$  & MAE $\downarrow$   & SR $\uparrow$   \\
                         & [dB]  & [bpm]   & [\%]  \\ \midrule
    Green~\cite{remoteAmbient}   &  -4.09 & 27.1 &15.0 \\
    ICA~\cite{PohICA}            &  -2.45 & 42.4 &8.8 \\
    CHROM~\cite{CHROM}           &  -6.56 & 38.2 &8.7 \\
    POS~\cite{POS}              &  1.66  & 59.6 &3.3 \\
    PVM~\cite{macwanPVM}         &  -6.82 & 72.5 & 6.3  \\
    Ours                         &  \textbf{2.67}  & \textbf{16.2} & \textbf{27.3}  \\ \bottomrule
\end{tabular}
\end{table}
 \begin{figure}[t]
    \centering
    \subfloat[]{\includegraphics[width=.9\linewidth]{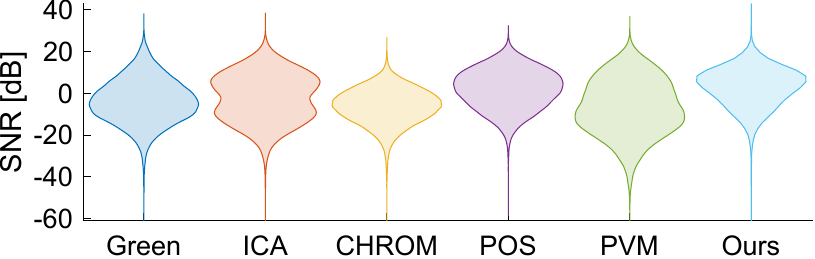}}\vspace{-2em}
    \subfloat[]{\includegraphics[width=.9\linewidth]{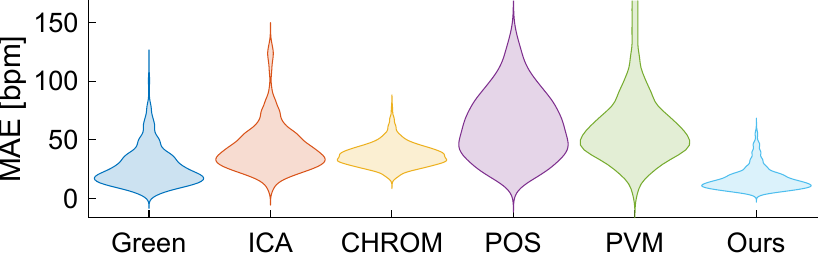}}\vspace{-2em}
    \subfloat[]{\includegraphics[width=.9\linewidth]{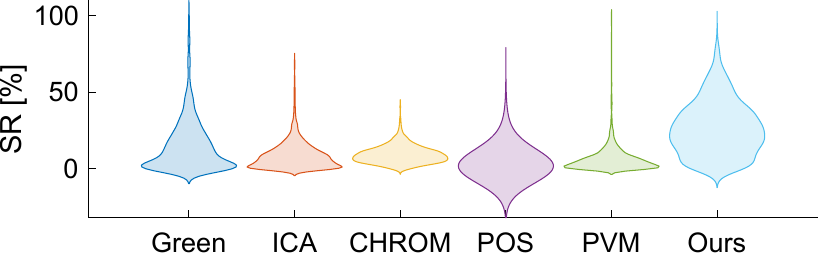}}\vspace{-1.5em}
    \caption{Violin plots of metrics for single patch setting.}
    \label{fig:violin}
 \end{figure}
 \begin{figure}[h]
    \includegraphics[width=0.95\linewidth]{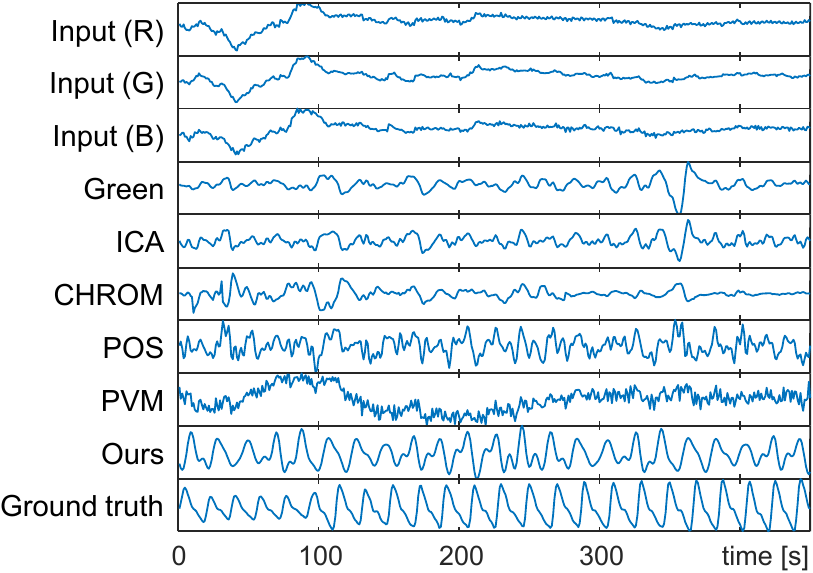}
    \caption{Example of input signals and extracted rPPG signals for single patch setting. (Subject \#47)}
    \label{fig:example}
 \end{figure}

\subsubsection{Parameter settings}
Through preliminary experiments, we set the parameters as follows.
For both the proposed and comparison methods, we set the number of input frames as $L=450$.
We set the window size $\tau$ and the smoothness parameter $\beta$ to 256 and 0.4, respectively.
We set the sparse parameters $\alpha_f = 0.1$ for $f \in [0.7, 4]~\mathrm{Hz}$.
This frequency range was set based on medical knowledge of cardiac cycles for normal persons~\cite{HR_range}.
For other frequency ranges, we set $\alpha_f = 100$.
We set the number of iterations for the alternating minimization to 15.
For ADMM parameters, we set $K = 20$, $\gamma = 1$.

\subsubsection{Evaluation metrics}
Following~\cite{CHROM, POS}, we quantitatively evaluated performance using the signal-to-noise ratio (SNR), defined as the ratio of the power spectral density at frequencies associated with rPPG to that at all other frequencies.
We determined the frequencies associated with rPPG by identifying the fundamental and second harmonic frequencies of the ground truth PPG signal.

In addition, using the extracted rPPG signal, we evaluated heart rate (HR) estimation performance, which is one of the most practical applications for rPPG measurement~\cite{niuTIP,ieice_ours}.
To compute HR, we analyzed the peak-to-peak intervals of the extracted rPPG signal and calculated HR in bpm.
As metrics, we computed the mean absolute error (MAE) and the success rate (SR) for the obtained HRs against the ground truth, following previous studies~\cite{ouraccess}.

\subsection{Experimental results}
We first report the results for the single patch setting.
Table~\ref{tab:results} shows the comparison results of the average performance of all patches.
Higher values indicate better performance for SNR and SR, whereas lower values indicate better performance for MAE.
It can be seen that the proposed method outperformed all comparison methods across all metrics.
Figure~\ref{fig:violin} shows the violin plots of each evaluation metric.
We can see that the metrics for the proposed method are distributed at higher performance levels than the comparison methods.
Figure~\ref{fig:example} shows an example of input RGB signals and the extracted rPPG signal, where each signal is normalized to zero mean and unit variance for visual clarity.
It can be seen that the proposed method was able to extract rPPG signals that closely resemble the ground truth PPG signals.

\begin{figure}[t]
    \centering
    \subfloat[]{\includegraphics[width=.9\linewidth]{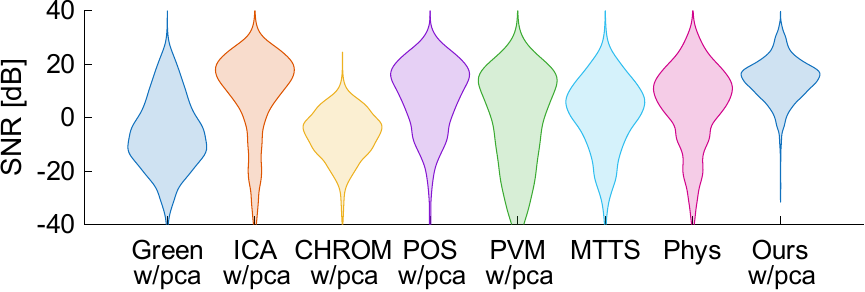}}\vspace{-2em}
    \subfloat[]{\includegraphics[width=.9\linewidth]{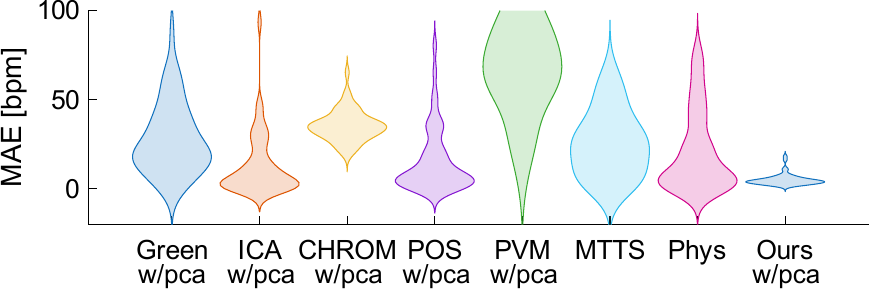}}\vspace{-2em}
    \subfloat[]{\includegraphics[width=.9\linewidth]{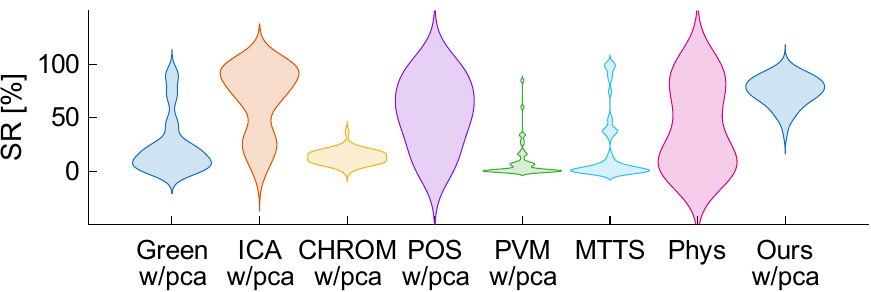}}\vspace{-1.5em}
    \caption{Violin plots of evaluation metrics for multiple patch setting.}
    \label{fig:violin-spatial}
\end{figure}

\begin{figure}[!h]
    \includegraphics[width=0.95\linewidth]{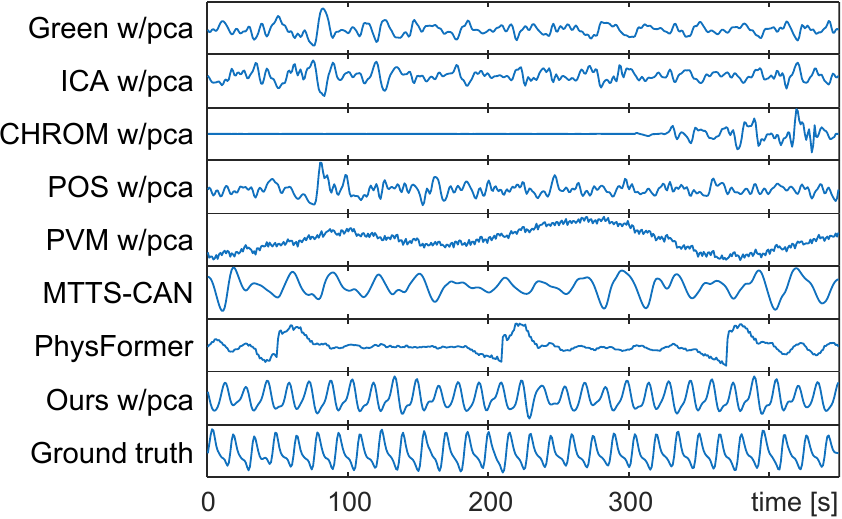}
    \caption{Example of rPPG signals for multiple patch setting (Subject \#15).}
    \label{fig:example-spatial}
 \end{figure}

For the multiple patch setting, we show the comparison results in Table~\ref{tab:results-spatial}.
The violin plots of each evaluation metric are shown in Fig.~\ref{fig:violin-spatial}.
We can see that the proposed method shows superior performance to comparison methods across all metrics.
We show an example of an extracted rPPG signal in Fig.~\ref{fig:example-spatial}, where each signal is normalized for visual clarity.
While \textit{PVM w/pca}, \textit{MTTS-CAN}, and \textit{PhysFormer} extract spurious periodic components, the proposed method extracts an rPPG signal that closely matches the ground truth signal.
This result demonstrates the effectiveness of our block-sparse model in the time-frequency domain.

 \begin{table}[t]
    \centering
    \caption{Performance comparison for multiple patch setting. The best results are highlighted in \textbf{bold}.}
    \label{tab:results-spatial}
\begin{tabular}{l|ccc} \toprule
    \multirow{2}{*}{Method} & SNR $\uparrow$  & MAE $\downarrow$   & SR $\uparrow$   \\
                         & [dB]  & [bpm]   & [\%]  \\ \midrule
    Green~\cite{remoteAmbient} w/pca      & -5.81 & 28.0 & 23.5 \\
    ICA~\cite{PohICA} w/pca        & 8.86  & 15.8 & 67.1 \\
    CHROM~\cite{CHROM} w/pca      & -6.29 & 35.4 & 13.2 \\
    POS~\cite{POS} w/pca        & 7.77  & 15.8 & 54.4 \\
    PVM~\cite{macwanPVM} w/pca        & -0.96 & 78.4 & 8.6  \\
    MTTS-CAN~\cite{mttscan}    & 0.71  & 25.4 & 21.1 \\
    PhysFormer~\cite{PhysFormer} & 4.13  & 21.3 & 41.3 \\
    Ours w/pca       & \textbf{13.98} & \textbf{5.1}  & \textbf{74.8}\\ \bottomrule
\end{tabular}
\end{table}

\section{Conclusion and Future Work}
\label{sec:conclusion}
We proposed an rPPG signal extraction method that exploits the quasi-periodic characteristics of rPPG signals.
We modeled the quasi-periodicity of rPPG signals as a block-sparse structure in the time-frequency domain.
To integrate our model into the rPPG signal extraction scheme, we constructed a time-varying signal separation framework.
This enables the incorporation of the block-sparse model as a signal prior while adaptively separating signals under fluctuating illumination.

For future work, we plan to integrate learning-based approaches into our framework.
For example, based on model-based deep learning frameworks such as deep unfolding~\cite{model-based-deep,otobeicip}, we will investigate the construction of neural networks that can directly incorporate our model-based algorithm.
This approach would enable performance improvement through supervised learning while preserving interpretability through our model-based formulation.

\bibliographystyle{IEEEbib}
\bibliography{reference}

\end{document}